\documentclass{article}
\usepackage{amsmath,graphicx}
\usepackage{booktabs}
\usepackage{xcolor}
\usepackage{caption} 
\usepackage{amsfonts}
\usepackage[preprint]{spconf}

\title{Invertible Voice Conversion}

\name{Zexin Cai$^{1}$, Ming Li\sthanks{Corresponding author: Ming Li}$^{1, 2}$ }
\address{$^{1}$Department of Electrical and Computer Engineering, Duke University, Durham, NC 27708, USA\\
        $^{2}$Data Science Research Center, Duke Kunshan University, Kunshan 215316, PR China}
\email{\{zexin.cai, ming.li369\}@duke.edu}
\begin{document}
\ninept

\maketitle

\begin{abstract}
In this paper, we propose an invertible deep learning framework called INVVC for voice conversion. It is designed against the possible threats that inherently come along with voice conversion systems. Specifically, we develop an invertible framework that makes the source identity traceable. The framework is built on a series of invertible $1\times1$ convolutions and flows consisting of affine coupling layers. We apply the proposed framework to one-to-one voice conversion and many-to-one conversion using parallel training data. Experimental results show that this approach yields impressive performance on voice conversion and, moreover, the converted results can be reversed back to the source inputs utilizing the same parameters as in forwarding. 
\end{abstract}
\section{Introduction}
\label{sec:intro}
Voice conversion (VC) offers the ability to transform the voice from a source audio signal to another desired voice while leaving the linguistic content unchanged \cite{sisman2020overview, mohammadi2017overview}. Over the past decade, the neural network-based VC \cite{laskar2012comparing, sun2015voice, fang2018high, zheng2016text} has surpassed traditional VC approaches \cite{abe1990voice, toda2007voice, zen2010continuous} in delivering natural synthesis results. It works well in both parallel and non-parallel training data conditions. When equipped with neural vocoders \cite{oord2016wavenet, Korte2020}, neural network-based VC systems can generate high-fidelity results that are significantly as natural as human speech. Such high-fidelity synthesized speech makes it difficult for people to distinguish whether it is real or not. On the other hand, the recent many-to-many VC approaches make it more powerful in voice cloning \cite{lin2021fragmentvc, chou19_interspeech}. Variable kinds of voices are available for conversion with the many-to-many setup. Moreover, the requirement of enrollment audio recording is further reduced by zero-shot conversion, which requires only seconds of audio from the target speaker for voice cloning \cite{qian2019autovc, zhang20e_interspeech, yang2021building}. 

Nonetheless, those aforementioned achievements unavoidably empower the spoofing attack on automatic speaker verification (ASV) systems \cite{kinnunen2012vulnerability}. Many voice services today rely on speaker verification, one of the biometric authentication means, to verify a person's identity.  Numerous studies demonstrate the vulnerabilities of speaker verification/recognition systems from spoofing attacks by voice conversion \cite{kinnunen2012vulnerability, wu2015spoofing}. Accordingly, there are works investigating countermeasures in response to those attacks. Developing more robust ASV systems and incorporating an independent spoofing detector are two popular countermeasures to overcome the threat, while the latter one gains more interest in this field \cite{wu2017asvspoof}. Subsequently, the biyearly ASVspoof challenge started in 2015 is held to encourage and facilitate researchers in terms of improving the spoofing detection performance \cite{wu2015asvspoof}. Tak et al. propose Rawnet2 as an anti-spoofing system in the challenge \cite{9414234}, which achieves impressive performance against voice conversion spoofing attack \cite{wang21_asvspoof}. 

However, many countermeasures against voice conversion from the spoofing literature aim to identify whether an audio signal is synthetic or not, while they are unable to trace the source of fraudsters. In this paper, we propose the INVVC, a bijective model for voice conversion, where the source speech can be obtained from the converted result by the same parameter set. We take advantages of invertible network modules, including $1\times1$ convolutions \cite{NEURIPS2018_d139db6a} and affine coupling layers \cite{DinhSB17} in our model. A variant transformer encoder \cite{vaswani2017attention} is used as the principal part to transform features in affine coupling layers. Our proposed model is trained and evaluated on a parallel dataset. Accordingly, we conduct experiments on one-to-one and many-to-one voice conversion using INVVC. The experimental results show that the proposed architecture achieves successful performance in both conversion cases. In particular, as an advantage of applying reversible networks, the source speech can be traced back from the converted results. 

This paper is organized as follows: In Section \ref{sec:bg}, we briefly review the background of invertible networks. In Section \ref{sec:med}, we present our proposed model and details on its components. Our experimental results are shown in Section \ref{sec:exp}, and we conclude our work in Section \ref{sec:con}.  

\begin{figure*}[ht]
  \centering
  \includegraphics[scale=0.28]{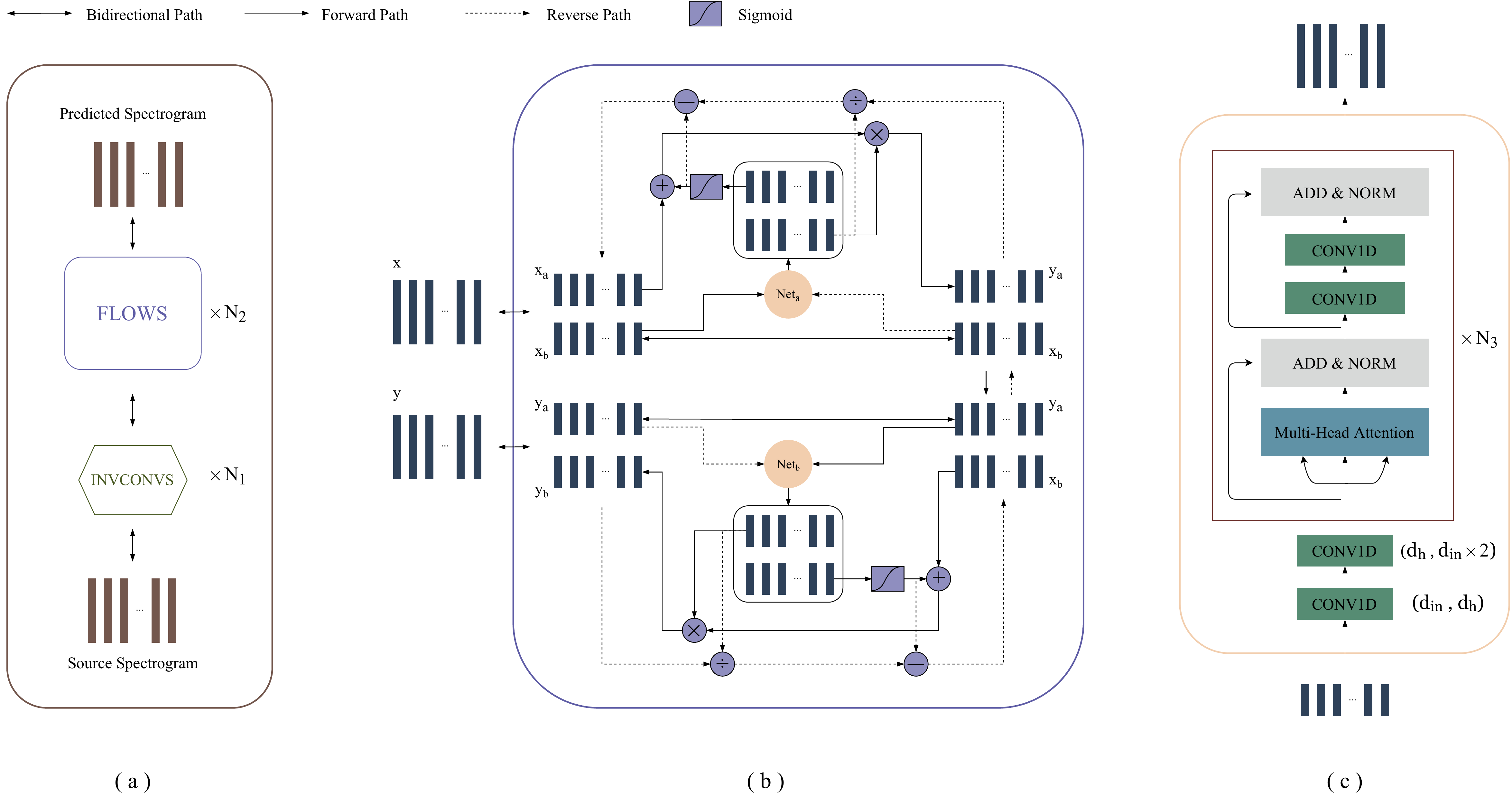}
  \caption{Proposed Inversible Voice Conversion (INVVC) Architecture}
  \label{fig:invvc}
\end{figure*}

\section{Background: Invertible Networks}
\label{sec:bg}
Invertible neural networks (INNs) have attracted great interest and achieved impressive results these years in different research fields, including image generation \cite{DinhSB17, NEURIPS2018_d139db6a} and speech synthesis \cite{8683143}. It is proposed for generative models to predict synthetic results from a standard probability distribution. Generally, invertible networks contain a series of bijective transformations called flow \cite{NICEDinhKB14}. Given high-dimensional vector $\mathbf{x} \in \mathbb{R}^d$, the network can be regarded as a bijective function $f_\theta$ that convert input $\mathbf{x}$ to output $\mathbf{y} \in \mathbb{R}^d$ from a desired distribution. In particular, the function $f_\theta$ can be broke down into K consecutive transformations as below,
$$\mathbf{y} = f_\theta(\mathbf{x})$$
$$f_\theta = f_1 \circ f_2 \circ \cdots \circ f_K$$
where each transformation is reversible. The above transformations are also known as forwards, while backwards are defined as the corresponding reverse computations that map output $\mathbf{y}$ back to $x$.
$$\mathbf{x} = f^{-1}_\theta(\mathbf{y})$$
$$f^{-1}_\theta = f_K^{-1} \circ \cdots \circ f_2^{-1} \circ  f_1^{-1}$$

Applying such reversible property of INNs in voice conversion, we are allowed to trace the source of a converted signal. Previously, INNs are used in many-to-many voice conversion called BLOW \cite{BLOWSerraPS19}, where target $\mathbf{y}$ is used to extract the conditional feature for transforming $\mathbf{x}$ to latent representation $\mathbf{z}$. Different from BLOW, we aim to design an invertible architecture that enables the bijective mapping between $\mathbf{x}$ and $\mathbf{y}$. Here $\mathbf{x}$ and $\mathbf{y}$ refer to the acoustic signals/features of the source speaker and the target speaker, respectively. In terms of likelihood-based generative models like BLOW, INNs are required to have tractable Jacobian determinants for the training criterion \cite{NICEDinhKB14}. However, we only take advantage of the reversible property of INNs in this paper. Thus our proposed model can be trained directly with criteria related to estimation quality, e.g., mean square loss.

\section{Proposed Framework: INVVC}
\label{sec:med}
We propose a novel architecture called INVVC for voice conversion. The overall conversion framework is shown in figure \ref{fig:invvc} (a), where the transformation from the source spectrogram to the target spectrogram is performed by a series of $1\times1$ invertible convolutions and FLOWs. The input spectrogram is the acoustic feature extracted from audio signals, like Mel-spectrogram. The number of invertible convolutions (INVCONVS) is $N_1$ and the number of FLOWs is $N_2$ as shown in the figure. While flow refers to a sequence of transformations in the literature \cite{NICEDinhKB14}, we denote a block with two consecutive affine coupling layers (see Section \ref{sec:afl}) as one step of FLOW here. The structure of FLOW is shown in figure \ref{fig:invvc} (b), while the nonlinear network component \textit{Net} is shown in Figure \ref{fig:invvc} (c).

\subsection{$1\times1$ Invertible Convolution}
As proposed in Glow \cite{NEURIPS2018_d139db6a}, $1\times1$ convolution is equivalent to an invertible permutation operation that mixes information across channels. We apply $1\times1$ convolution in our conversion framework to reduce the restriction on information exchange among channels. The weight $\mathbf{W}$ of the $1\times1$ convolution is a square matrix initialized to be orthonormal. Therefore the matrix $\mathbf{W}$ is invertible. The forward operation of $1\times1$ convolutions is 
$$\forall i: \mathbf{y}_i = \mathbf{W}\mathbf{x}_i,$$
and the reverse operation is 
$$\forall i: \mathbf{x}_i = \mathbf{W}^{-1}\mathbf{y}_i,$$,
where $i$ denotes the frame indices into feature map $\mathbf{x}$ and $\mathbf{y}$.

\subsection{FLOW}
\label{sec:afl}
A step of FLOW is an alternating pattern consisting of two consecutive affine coupling layers. As shown in figure \ref{fig:invvc} (b), a FLOW can be separated to the upper part that leaves half of the input channels $\mathbf{x}_b$ unchanged, and the symmetrical lower part that transforms $\mathbf{x}_b$. Particularly, the affine coupling layer is a powerful bijective operation that we use to transform the source feature to the target feature \cite{DinhSB17}. As one affine coupling layer needs to leave some value unchanged to maintain invertibility, we apply it in such an alternating manner to fully transform the input. The forward and reverse operations of the upper coupling layer can be formulated as below:

\linespread{1.1}
\begin{table}[h]
\small
\centering
\begin{tabular}{c|c}
 \footnotesize{FORWARD} & \footnotesize{REVERSE}  \\ \hline
 $\mathbf{x}_a, \mathbf{x}_b = \textsc{split}(\mathbf{x})$ &  $\mathbf{y}_a, \mathbf{y}_b^{\prime} = \textsc{split}(\mathbf{y}^{\prime})$ \\
 $(\mathbf{u}, \mathbf{t}) = \textit{Net}_a(\mathbf{x}_b)$ &  $(\mathbf{u}, \mathbf{t}) = \textit{Net}_a(\mathbf{y}_b^{\prime})$ \\
 $\mathbf{s} = \sigma(\mathbf{u} + \epsilon)$ &  $\mathbf{s} = \sigma(\mathbf{u} + \epsilon)$ \\
 $\mathbf{y}_a = \mathbf{s} \odot \mathbf{x}_a + \mathbf{t}$ & $\mathbf{x}_a = (\mathbf{y}_a - \mathbf{t}) / \mathbf{s}$ \\
 $\mathbf{y}_b^{\prime} = \mathbf{x}_b$ & $\mathbf{x}_b = \mathbf{y}_b^{\prime}$ \\
 $\mathbf{y}^{\prime} = \textsc{concat}(\mathbf{y}_a, \mathbf{y}_b^{\prime})$ &  $\mathbf{x} = \textsc{concat}(\mathbf{x}_a, \mathbf{x}_b)$
\end{tabular}
\end{table}

where $\textsc{split}()$ is the operation that chunks the input feature into two halves along the channel dimension, $\textit{Net}_a$ is a nonlinear mapping structure that increases the channel dimension so that we can obtain intermediate variables $\mathbf{u}$ and $\mathbf{t}$ with appropriate size. $\mathbf{s}$ is the element-wise scale vector, $\epsilon$ is a constant, and $\sigma$ is the sigmoid function. The forward and reverse operations of the bottom coupling layer is similar, while it keeps $\mathbf{y}_a$ unchanged and convert $\mathbf{y}_b^\prime$, which is $\mathbf{x}_b$, to $\mathbf{y}_b$.

\subsection{Conversion \textit{Net}}
Generally, the component \textit{Net} above can be any network structure that doubles the dimensionality of the input feature. \textit{Net} helps obtain the scaling factor $u$ and the affine component $t$ for feature conversion. Here we adopt a transformer-based structure for transforming the feature \cite{vaswani2017attention}. As shown in figure \ref{fig:invvc} (c), we use two 1D convolutions to pre-encode the input first, where the second one doubles the channel dimension. Then $N_3$ identical blocks are followed after the pre-encoding layers. Each block has a multi-head attention module and a convolution module with two 1D convolutional layers. Both parts have residual connections, followed by layer normalization. 

\section{Experiments}
\label{sec:exp}
We investigate the performance of our proposed invertible model by one-to-one voice conversion and many-to-one conversion. Accordingly, we put audio samples online for readers to listen to. \footnote{https://caizexin.github.io/invvc/index.html}.


We choose the CMU ARCTIC \cite{kominek2003cmu} English corpus as our primary dataset for experiments. Seven speakers\footnote{ http://festvox.org/cmu\_arctic/} are chosen from the CMU ARCTIC database in the following experiments, which are notated as `bdl', `slt', `clb', `rms', `jmk', `awb' and `ksp'. This phonetically balanced dataset contains 1132 parallel speech utterances for each speaker. All utterances are distributed in 16KHz waveforms. For each speaker, 1000 utterances are used for training, while the remaining 132 utterances are used as the test set. 

\subsection{Training Setup}
\label{sec:train}
The Mel-spectrogram is used as the acoustic feature in our experiment. 80-dim Mel-spectrograms are extracted with a 25ms window length every 12.5ms from the continuous speech. We use Dynamic Time Warping (DTW) to align the acoustic features for parallel pairs during training. 

Regarding the hyperparameters of the network, the number of the invertible $1\times1$ convolution $N_1$ is set to 2, while the number of the transformer-based encoder block $N_3$ is set to 4 in all following experiments. In particular, we use 4 FLOWs in one-to-one conversion and use 6 FLOWs in the many-to-one scenario. As for the conversion \textit{Net}, the kernel size of two pre-encoding convolutional layers is set to 3, while the channel size $d_h$ is set to 512. We use two heads for the multi-head attention mechanism. The two convolutional layers inside the transformer-based block are similar to the pre-encoding layers. However, the in-between channel size is set to 1032, and filter sizes for the two convolutions are 9 and 1, respectively.

The training is optimized with respect to the mean square error (MSE) between the predicted Mel-spectrogram and the ground-truth spectrogram. Other than the MSE loss, we have the mean absolute error (L1) loss between the means of the two Mel-spectrograms as an additional criterion. Similarly, the L1 loss between standard deviations of two Mel-spectrograms is also added in our training.  The final training loss can be formulated as in equation \ref{eqa:loss}. 

\begin{equation}
\label{eqa:loss}
    \begin{split}
        L_\text{train} &= \text{MSE}(\text{Predicted\_Mel}, \text{Target\_Mel}) \\
        & + L1(mean(\text{Predicted\_Mel}), mean(\text{Target\_Mel})) \\
        & + L1(std(\text{Predicted\_Mel}), std(\text{Target\_Mel}))
    \end{split}
\end{equation}

We use Adam optimizer with an initial learning rate set to 0.0001 and beta set to (0.9, 0.98). The neural vocoder WaveRNN \cite{kalchbrenner2018efficient} is used to convert the acoustic feature back to audio signals. The vocoder is trained with the training set of the seven speakers. In this case, there are approximately 6.4 hours, with 7000 utterances, in total for vocoder training. 


\subsection{One-to-one Conversion}
We first study our proposed model on one-to-one voice conversion, where we aim to convert bdl's voice to slt's voice. In this case, we set the number of FLOWs $N_2$ to 4. For each utterance from the test set, we infer the corresponding speech of `bdl' to the trained model. The voice conversion results are in slt's voice, which is annotated as VC. Then we reverse the converted results back and annotated those inverted results as INV, which can be seen as the simulation of tracing the source of the voice conversion. All results are converted to waveforms by the neural vocoder described in section \ref{sec:train}. 

We randomly select 20 ground truth natural utterances (notated as GT below), 20 converted ones from VC, and 20 inverted ones from INV for subjective evaluation. 20 participants are asked to rate these utterances regarding naturalness and similarity. The scores are in MOS-scale, a categorical score from 1 to 5. Here similarity rating is carried out by comparing the voice of the synthesized utterance to the voice of a natural utterance from the expected speaker. For instance, inverted results are rated with respect to natural utterances from source speaker `bdl', while the speech contents of each comparing pair are different. 

\begin{table}[htbp] 
 \caption{MOS results on one-to-one conversion} 
 \label{tbl:o2o}
  \centering  
 \begin{tabular}{cccc}    
 \toprule   
  MOS $\pm$ 95\% CI   & GT  & VC & INV  \\
 \midrule  
  Naturalness & 4.57$\pm$0.05 & 3.76$\pm$0.08 & 3.74$\pm$0.08  \\
  Similarity & - & 4.2$\pm$0.06 & 4.14$\pm$0.07  \\
\bottomrule 
\end{tabular}  
\end{table}

MOS results, along with 95\% confidence intervals, are shown in table \ref{tbl:o2o}. For naturalness, the converted results of our proposed model achieve an average score of around 3.76, which is close to the inverted results. The naturalness of the natural speech is around 4.57. Since our network is invertible, the inverted result is the same as those utterances from the source speaker `bdl'. However, the quality downgrades slightly due to the waveform reconstruction loss from the vocoder. Therefore, the score of INV can also be seen as the synthesis performance of the vocoder. Concerning similarity, results show that the model achieves a score of 4.2, which implies that the voice is close to our desired voice. As can be seen, our proposed model achieves satisfactory results on the one-to-one conversion scenario.

\subsection{Many-to-one Conversion}
Likewise, we investigate the performance of our model on converting multiple voices to the same voice. $N_2$ is set to 6 in this scenario. The speaker `bdl' is used as the target voice, and the remaining 6 speakers are treated as source speakers. We obtain conversion results (VC) and inverted results (INV) based on test utterances, similar to the one-to-one scenario. 

\begin{table}[htbp] 
 \caption{MOS results on many-to-one conversion} 
 \label{tbl:m2o}
  \centering  
 \begin{tabular}{cccc}    
 \toprule   
  MOS $\pm$ 95\% CI   & GT  & VC & INV  \\
 \midrule  
  naturalness & 4.59$\pm$0.05 & 3.65$\pm$0.07 & 3.99$\pm$0.07   \\
  similarity & - & 4.16$\pm$0.05 & 4.3$\pm$0.06  \\
\bottomrule 
\end{tabular}  
\end{table}

\noindent\textbf{Subjective evaluation.} Again, 21 ground truth utterances, with 3 from each speaker (including the target speaker), are chosen for evaluation. 24 converted utterances, with 4 regarding each source speaker, and the same from inverted results, are randomly selected for subjective evaluation. The MOS results on naturalness and similarity are shown in table \ref{tbl:m2o}. The overall naturalness score of the converted utterances is 3.65, while the inverted results achieve a score of 3.99. Similar to the one-to-one case, the score of INV indicates the synthesis performance of the vocoder. Comparing with the naturalness results from table \ref{tbl:o2o}, we know that the naturalness score of bdl's voice, which is a male voice, is below average, while the score of slt's utterances, a female voice, is likely to be above average. Therefore, as we choose `bdl' as the target speaker in the many-to-many scenario, the naturalness score on conversion results is slightly lower. 

Nonetheless, the model's performance on similarity is impressive. The speaker similarity score of the converted results is around 4.16. This outcome indicates that the many-to-one conversion system trained with our proposed model can transform the voice accurately. Moreover, the score of inverted results is around 4.3, which shows that we are able to trace the source speaker from the reversed results when all source speakers' utterances are converted to the same voice.   

\noindent\textbf{Objective evaluation.} We evaluate the performance of our many-to-one system by two objective evaluation metrics. The first evaluation method is Mel-Spectrogram Distortion (MSD), using the same distance measurement of Mel Cepstral distortion (MCD) but different feature \cite{kubichek1993mel}. MSD measures the difference between the synthesized Mel-spectrogram sequence and the corresponding natural Mel-spectrogram sequence based on DTW. The MSD scores are calculated on the test set, and the results are shown in table \ref{tbl:m2o_obj}. The MSD score between parallel utterance pairs from source speakers and the target speaker is 12.42. However, the distortion rate decrease to 8.35 after the conversion. Furthermore, the invertibility of our proposed model can be verified by the MSD score between the inverted results and the source Mel-spectrograms. The MSD score, in that case, is close to 0, while the error mainly comes from the precision loss of the invertible convolutions.

\begin{table}[htbp] 
 \caption{Objective performance of the many-to-one model} 
 \label{tbl:m2o_obj}
  \centering  
 \begin{tabular}{cccc}    
 \toprule   
     & VC vs. Tgt  & Src vs. Tgt & Src vs. INV  \\
 \midrule  
  MSD\ (dB) & 8.35 & 12.42 & 0.03   \\
\bottomrule 
\end{tabular}  
\end{table}

\begin{figure}[ht]
  \centering
  \includegraphics[scale=0.24]{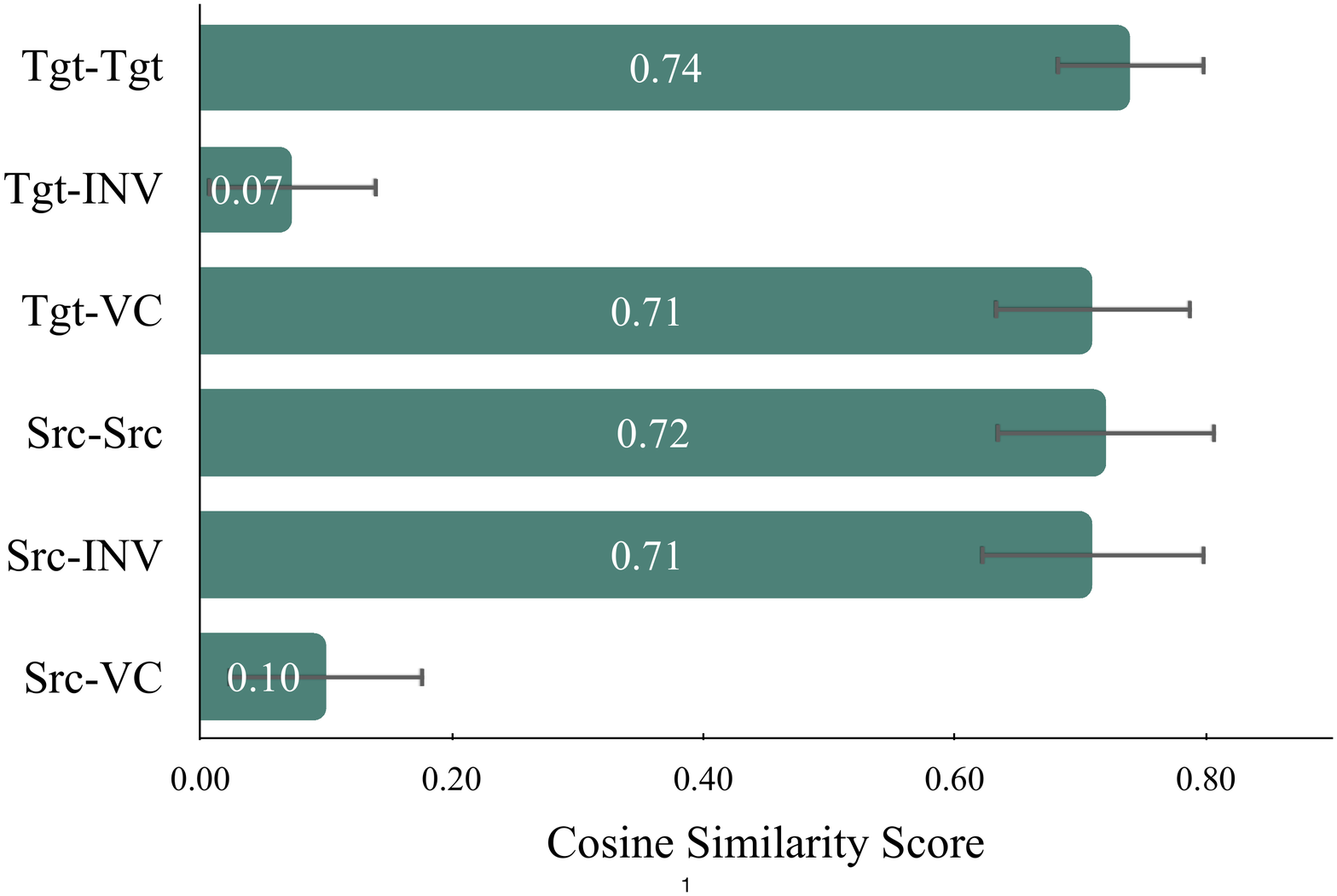}
  \caption{Objective performance on speaker similarity}
  \label{fig:cos}
\end{figure}

The other objective evaluation metric is for the speaker similarity. We use a pre-trained speaker verification model called ECAPA-TDNN \cite{desplanques2020ecapa} to extract speaker embeddings and calculate the cosine similarity scores between embedding pairs. The verification model achieves an Equal Error Rate (EER) of 0.4\% on our test set, implying its outstanding performance in distinguishing different voices. The similarity results are shown in figure \ref{fig:cos}. For each type of comparing set, we randomly create 1200 unique pairs for calculating the similarity scores. The `Tgt-Tgt' set measures the average cosine similarity score between two different utterances from the target speaker `bdl'. The `Tgt-INV` evaluates the similarity between inverted results and the natural utterances from the target speaker. The `Tgt-VC` contains pairs between converted utterances and natural utterances, while the `Src-Src' contains pairs between utterances from the same source speakers, with 200 utterance pairs from each speaker. The `Src-INV' evaluates the similarity between inverted results and the natural utterances, and the `Src-VC` presents the similarity of the converted utterances and their corresponding source utterances. As shown in the figure, the similarity score of `Tgt-VC' is close to the one of `Tgt-Tgt', which implies that the conversion voice of our model is pretty close to the desired voice from the verification system's perspective. In addition, we can see that the score of `Src-INV' is close to the score of `Src-Src'. Both sets have a high similarity score above 0.7. Again, this shows the ability of our invertible model to trace the source voice.

\setlength{\abovecaptionskip}{0pt plus 0pt minus 0pt}

\noindent\textbf{Unseen speakers.} Interestingly, models from both scenarios are available for converting unseen speakers' voices to the target speaker we trained on. However, we observe a distinguishable downgrade in the conversion quality, while the inversion quality stays the same. We encourage readers to listen to audio samples on our demo website for the conversion results on unseen speakers. 


\section{Conclusion}
\label{sec:con}
We proposed a bijective architecture for voice conversion while still capable of tracing the source speakers. By taking the advantages of invertible $1\times1$ convolutions and affine coupling layers, the model can reverse the conversion output back to the input feature. The model's performance is evaluated by experiments including one-to-one voice conversion and many-to-one conversion. Subjective evaluation results from both experiments indicate the impressive performance of our proposed model on voice conversion. Furthermore, the model's invertibility is demonstrated by elaborate objective measurements, where the inverted mel-spectrograms are almost the same as the original ones from the source inputs. Therefore, we can trace the source speaker from the converted speech with our proposed model. 

\bibliographystyle{IEEEbib}
{\footnotesize
\bibliography{refs}}

\end{document}